# Sub-1-μs, Sub-20-nJ Pattern Classification in a Mixed-Signal Circuit Based on Embedded 180-nm Floating-Gate Memory Cell Arrays

F. Merrikh Bayat[1], X. Guo[1], M. Klachko[1], M. Prezioso[1], K. K. Likharev[2], and D. B. Strukov[1]
[1]UC Santa Barbara, Santa Barbara, CA 93106-9560, U.S.A., [2]Stony Brook University, Stony Brook, NY 11794-3800, U.S.A.

***Abstract*****-**We have designed, fabricated, and successfully tested a prototype mixed-signal, 28×28-binary-input, 10-ouput, 3-layer neuromorphic network ("MLP perceptron"). It is based on embedded nonvolatile floating-gate cell arrays redesigned from a commercial 180-nm NOR flash memory. The arrays allow precise (~1%) individual tuning of all memory cells, having long-term analog-level retention and low noise. Each array performs a very fast (sub-μs) and energy-efficient analog vector-by-matrix multiplication, which is the bottleneck for signal propagation in most neuromorphic networks. All functional components of the prototype circuit, including 2 synaptic arrays with 101,780 floating-gate synaptic cells, 74 analog neurons, and the peripheral circuitry for weight adjustment and I/O operations, have a total area below 1 $mm^2$. Its testing on the common MNIST benchmark set (at this stage, with a relatively low weight import precision) has shown a classification fidelity of 94.65%, close to the 96.2% obtained in simulation. The classification of one pattern takes less than 1 μs time and ~20 nJ energy – both numbers much better than for digital implementations of the same task. Estimates show that this performance may be further improved using a better neuron design and a more advanced memory technology, leading to a $>10^2\times$ advantage in speed and a $>10^4\times$ advantage in energy efficiency over the state-of-the-art purely digital (GPU and custom) circuits, at classification of large, complex patterns.

## I. Introduction

The idea of using nonvolatile floating-gate memory cells in analog and mixed-signal artificial neural network circuits has been around for almost three decades [1]. Up until recently, most of its implementations were based on "synaptic transistors" [2], which may be fabricated using the standard CMOS technology. Some sophisticated, energy-efficient systems have been demonstrated [3, 4] using this approach. However, the synaptic transistors have relatively large areas ($\sim 10^3 F^2$, where $F$ is the minimum feature size), leading to larger time delays and energy consumption [5].

Fortunately, during the last decade the nonvolatile floating-gate memory cells have not only been highly optimized and scaled down all the way to $F \sim 20$ nm, but also may now be imbedded in CMOS integrated circuits [6]. These cells are quite suitable to serve as adjustable synapses in neuromorphic networks, provided that the memory arrays are redesigned to allow for individual, precise adjustment of the memory state of each device. Recently, such modification was performed [7, 8] using the 180-nm ESF1 embedded commercial NOR flash memory technology of SST Inc. [6] (Fig. 1), and, more recently, the 55-nm ESF3 technology of the same company, with good prospects for its scaling down to at least $F = 28$ nm. Though such modification nearly triples the cell area, it is still at least an order of magnitude smaller, in terms of $F^2$, than that of synaptic transistors [2-5].

The main result reported in this paper is the first successful use of this approach for the experimental implementation of a (so far, relatively simple) neuromorphic network, which can perform a high-fidelity classification of images of the standard MNIST benchmark, with record-breaking speed, density, and energy efficiency.

## II. Cell Characterization for Analog Operation

Our network design (see below) uses energy-saving gate coupling [1, 5] of the peripheral and array cells, which works well in the subthreshold mode, with a nearly exponential dependence of the drain current $I_{DS}$ of the memory cell on the gate voltage $V_{GS}$. Fig. 2 shows the dependences $I_{DS}(V_{GS})$ for several memory states of a 180-nm ESF1 NOR memory cell. With the requirement to keep the relative current fluctuations (Fig. 3b) below 1%, the dynamic range of the subthreshold operation is about five orders of magnitude, from ~10 pA to ~300 nA, with a gate voltage swing up to 1.5 V, depending on the memory state. As the inset in Fig. 2 shows, the subthreshold slope stays relatively constant for low conductance memory states, becoming more steep at small gate voltages, apparently due to the cell's split-gate design.

The ESF1 flash technology guarantees a 10-year digital-mode retention at temperatures up to 125°C [6]. Fig. 3 shows that these cells feature at least a-few-days analog-level retention, with very low fluctuations of the output current. Other features of the ESF1 cell arrays, including the details of their modification, switching kinetics and its statistics, and an experimental demonstration of a fast weight tuning with a ~0.3% accuracy, were reported in Refs. [7, 8].

## III. Network Design

The implemented neuromorphic network (Figs. 4, 5) is a 3-layer (one-hidden-layer) MLP perceptron with 784 inputs, representing 28×28 black-and-white pixels of the input patterns (such as shown in Fig. 4a), 64 hidden layer neurons, and 10 output neurons (Fig. 4b). Each neuron gets an additional input from a bias node. The hidden layer neurons implement a rectified-tanh activation function (Figs. 4b, 5b).

The synaptic coupling between the neuron layers is provided by two crossbar arrays of the floating-gate memory cells. With the differential-pair implementation of each synapse [7], the total number of utilized floating-gate cells is 2×[(28×28+1)×64 + (64+1)×10] = 101,780. The desirable synaptic weights are imported into the network by analog tuning of the memory state of each floating-gate cell, using special peripheral circuitry (Fig. 5a). For simplicity of this first, prototype design, weights were tuned one-by-one, by applying proper bias voltage sequences to selected and half-selected lines [7]. The input pattern bits are shifted serially into a 785-bit register before each classification; to start it, the bits are read out in parallel into the network.

The vector-by-matrix multiplication in the first crossbar array is implemented by applying input voltages (either 4.2 V or 0 V) directly to the gates of the array cell transistors (Fig. 5b). The resulting source currents, each equal to the sum of the binary voltage inputs multiplied by the pre-imported analog weights, feed operational amplifier pairs (Fig. 5e). Signals from two differential cell rows are subtracted by feeding the output of one opamp of the pair to the input of its counterpart (Fig. 5d), with the output passed to the activation function circuit (Fig. 5f). The fully analog vector-by-matrix calculation in the second array is implemented using the gate-coupled approach [1, 5] (Fig. 3b). To minimize the error due to the subthreshold slope's dependence on the memory state (Fig. 2), we use a higher gate voltage range (1.1 V to 2.7 V), limited only by technology restrictions. The 10 voltage outputs of the network are measured externally.

## IV. Experimental Results

The synaptic weights were calculated in an external computer using the standard error backpropagation algorithm, and then "imported" into the mixed-signal circuit, i.e. used to tune the memory state of each array cell to the proper value. In the reported first experiments, only ~30% of the cells were tuned, and the weight import accuracy for a single cell tuning was limited to 5%, to decrease the import time. As Fig. 7 indicates, some of the already tuned cells were disturbed beyond the target accuracy during the subsequent weight import. At this preliminary testing stage, these cells were not re-tuned - in part because even for such rather crude weight import, the experimentally tested classification fidelity (94.65%) on MNIST benchmark test patterns is already remarkably close to the simulated value (96.2%) for the same network (Fig. 8), and not too far from the maximum fidelity (97.7%) of a MLP perceptron of this size.

Excitingly, such classification fidelity is achieved at an ultralow (sub-20-nJ) energy consumption per average classified pattern (Fig. 9a), and the average classification time below 1 μs (Fig. 9b). (The upper bound of the energy is calculated as a product of the measured average power, 5.6 mA × 2.7 V + 2.9 mA × 1.05 V ≈ 20 mW, consumed by the network, by the upper bound, 1 μs, of the average signal propagation delay. A more accurate measurement of both the time delay and energy would require a redesign of the signal input circuit, currently rather slow – see Fig. 9b.)

## V. Discussion and Summary

The achieved speed and energy efficiency are much better than those demonstrated, for the same task, at any digital network we are aware of. These are also many ready reserves in the circuit design. For example, calculations show that current mirrors in neuron circuits may strongly decrease their currently dominant contribution to latency and energy (Fig. 9). Our estimates (Fig. 10) show that using such mirrors, and a more advanced 55-nm memory technology ESF3 of the same company [6] may provide an at least ~100× advantage in the operation speed, and an enormous, $>10^4\times$ advantage in the energy efficiency, over the state-of-the-art purely digital (GPU and custom) circuits [9], at classification of large, complex patterns using deep-learning convolutional networks – see, e.g., Ref. [10].

To summarize, even the preliminary results reported here give an important demonstration of the exciting possibilities opened for neuromorphic networks by mixed-signal circuits based on industrial-grade floating-gate memory technologies.

## Acknowledgment

This work was supported by DARPA's UPSIDE program under contract HR0011-13-C-0051UPSIDE via BAE Systems. The authors are grateful to P.-A. Auroux, M. Bavandpour, N. Do, J. Edwards, M. Graziano, and M. R. Mahmoodi for useful discussions and technical support.

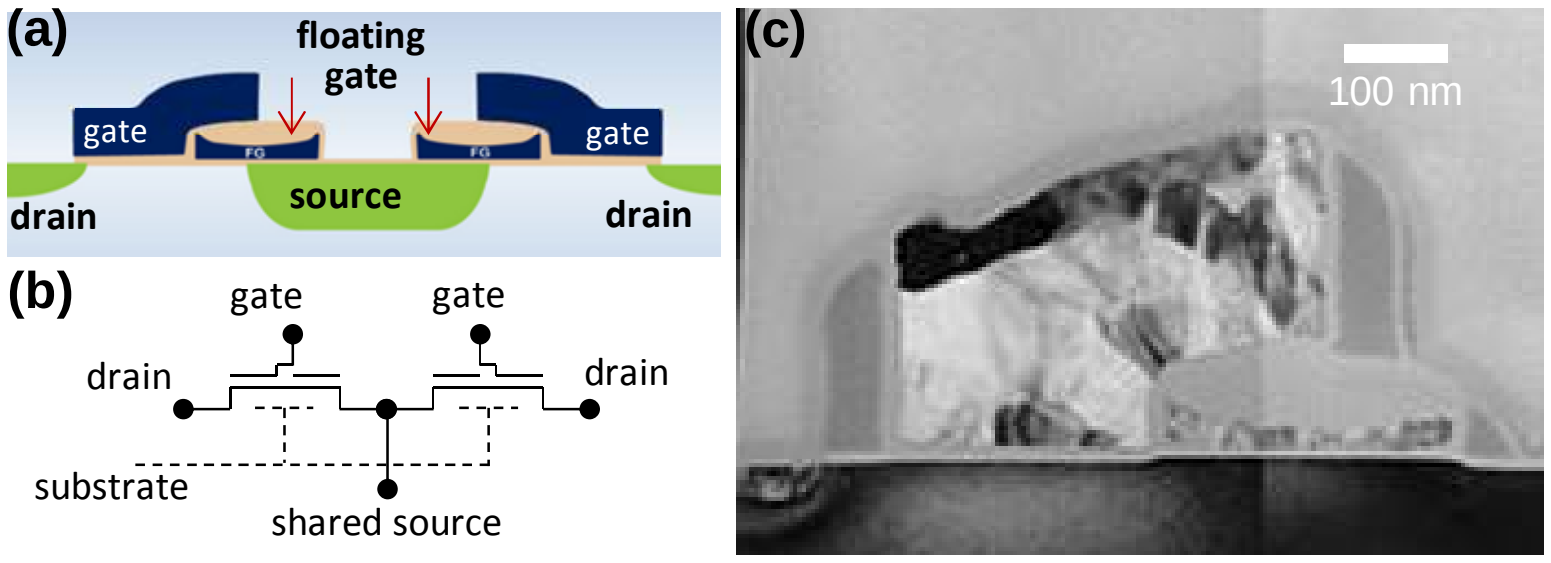

Fig. 1. SST's ESF1 NOR flash memory cells: (a) Cross-section of the two-cell "supercell" (schematically), (b) its equivalent circuit, and (c) TEM cross-section image of one memory cell in a 180 nm process.

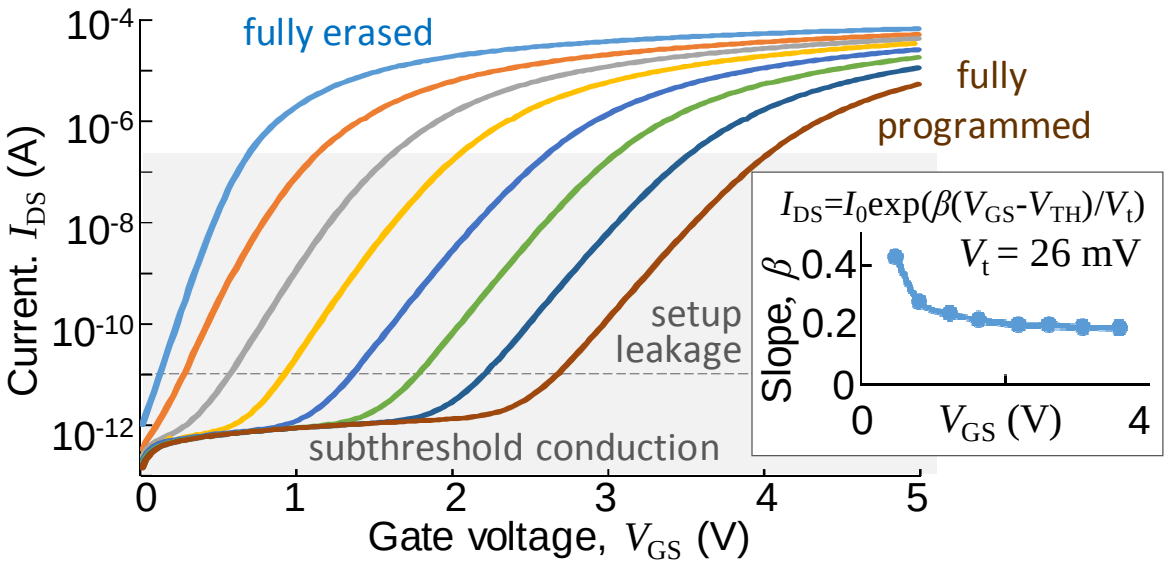

Fig. 2. Drain current of a ESF1 cell, at $V_{DS}$ = 1 V, as a function of the gate voltage, for several memory states. Inset: the log slope $\beta$, measured at $I_{DS}$ = 10 nA, as a function of the memory state (shown as a corresponding gate voltage).

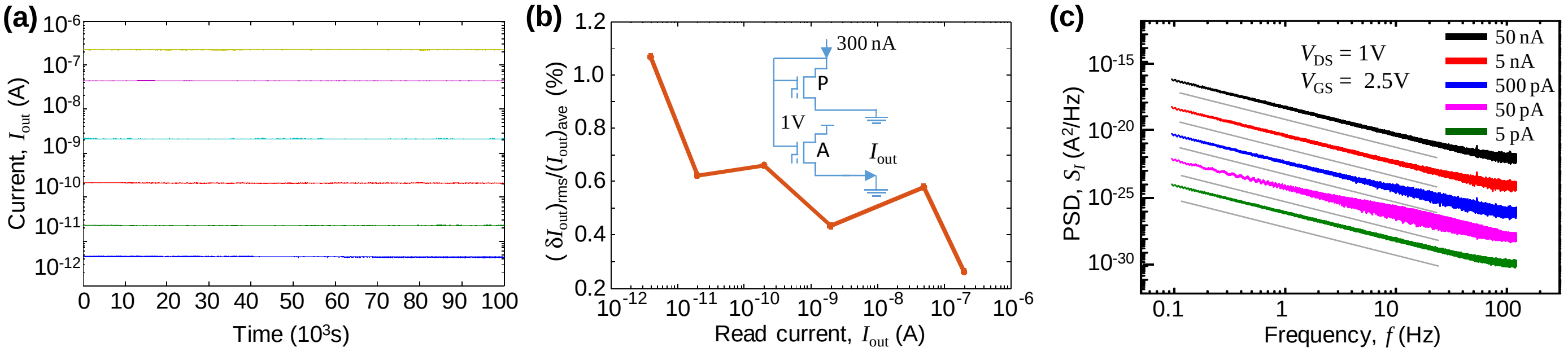

Fig. 3. (a) Retention measurements for several memory states, performed in the gate-coupled array configuration, and (b) the average relative variation of the currents during the same time interval. The inset shows the equivalent circuit of the used gate coupling. Each point on panel (a) is an average over 65 samples taken within a 130 ms period. (c) Spectral density of cell current's noise measured at room temperature; the gray lines are just guides for the eye, corresponding to $S_I \propto 1/f^{1.6}$.

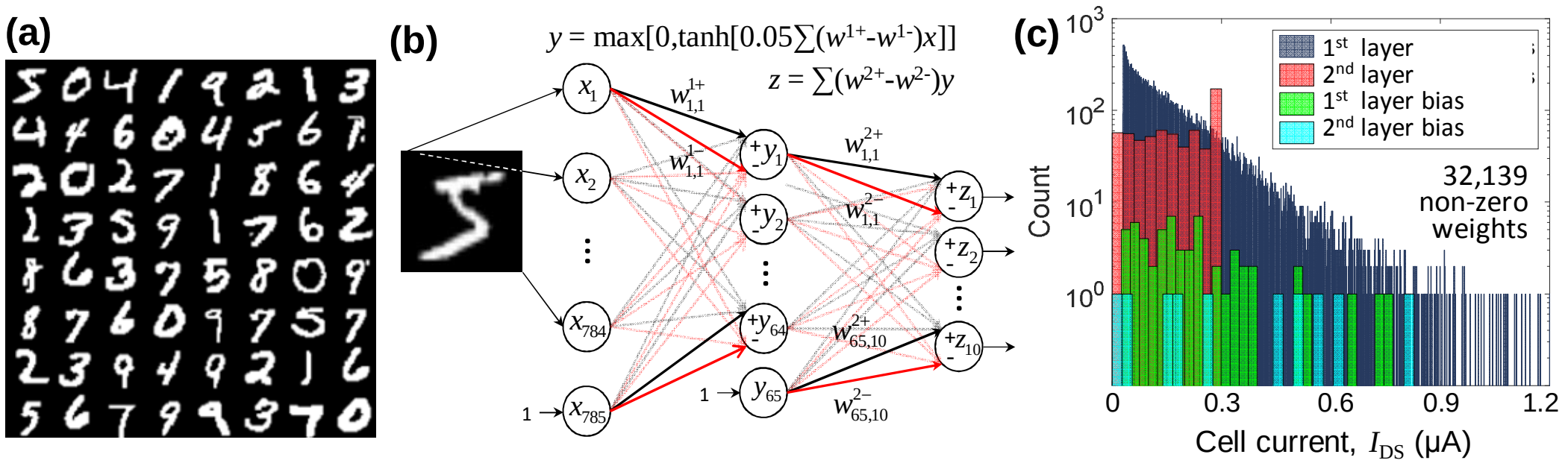

Fig.4. (a) Typical examples of B/W hand-written digits in the MNIST benchmark. (b) Graph representation of our 3-layer perceptron network. Each synapse is implemented using a differential pair of floating-gate memory cells. (c) A histogram showing the imported weight / cell current values (at $V_D$ = 2.7 V, and $V_S$ =1.65 V and $V_G$ = 4.2 V in the 1st array, and $V_S$ = 1.1 V and $V_G$ =2.7 V for the 2nd one), which were used in the experiment.

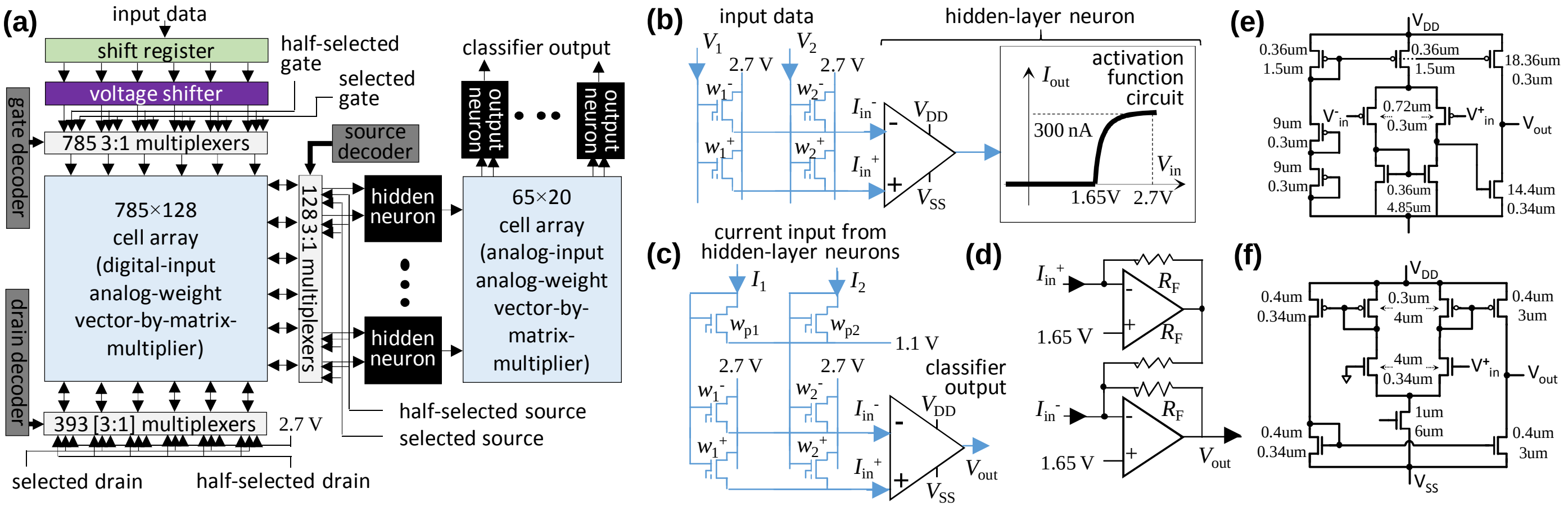

Fig.5. The network: (a) High-level architecture, with the weight tuning circuitry for the second array (similar to that of the first one) not shown for clarity. The voltage shifter enables using voltage inputs of both polarities over a 1.65V bias, and is also used to initiate the classification process by increasing the input background from 1.8 V to 4.2 V. (b) A 2×2-cell slice of the first crossbar array shown together with a hidden-layer neuron, which consists of a differential summing opamp pair and an activation-function circuit. (c) A 2×2-cell slice of the second crossbar array with an output-layer neuron; these neurons do not implement an activation function. (d) Circuit-level diagram of a differential summing amplifier used in the hidden-layer and output-layer neurons; $R_F$ =16 KΩ for hidden, and $R_F$ =128 KΩ for output neurons. (e, f) Transistor-level schematics of: (e) the operational amplifier and (f) the activation function; $V_{SS}$ = 0V, $V_{DD}$ = 2.7 V.

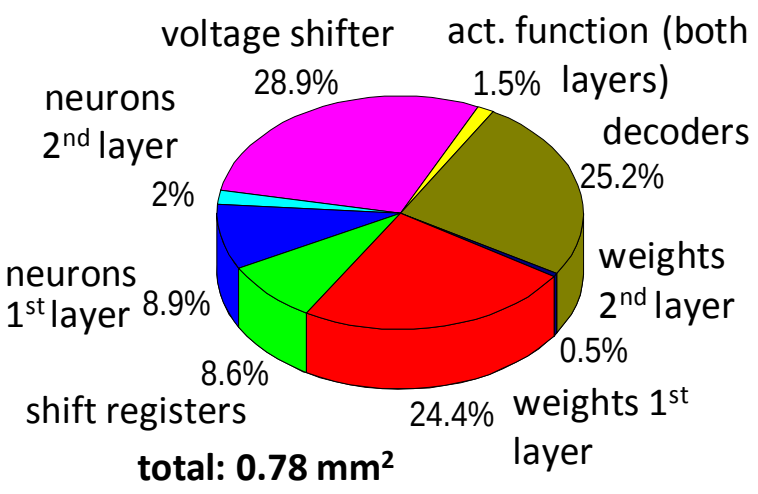

Fig. 6. Active component area breakdown (without accounting for wiring between the blocks, which is not optimized at this stage).

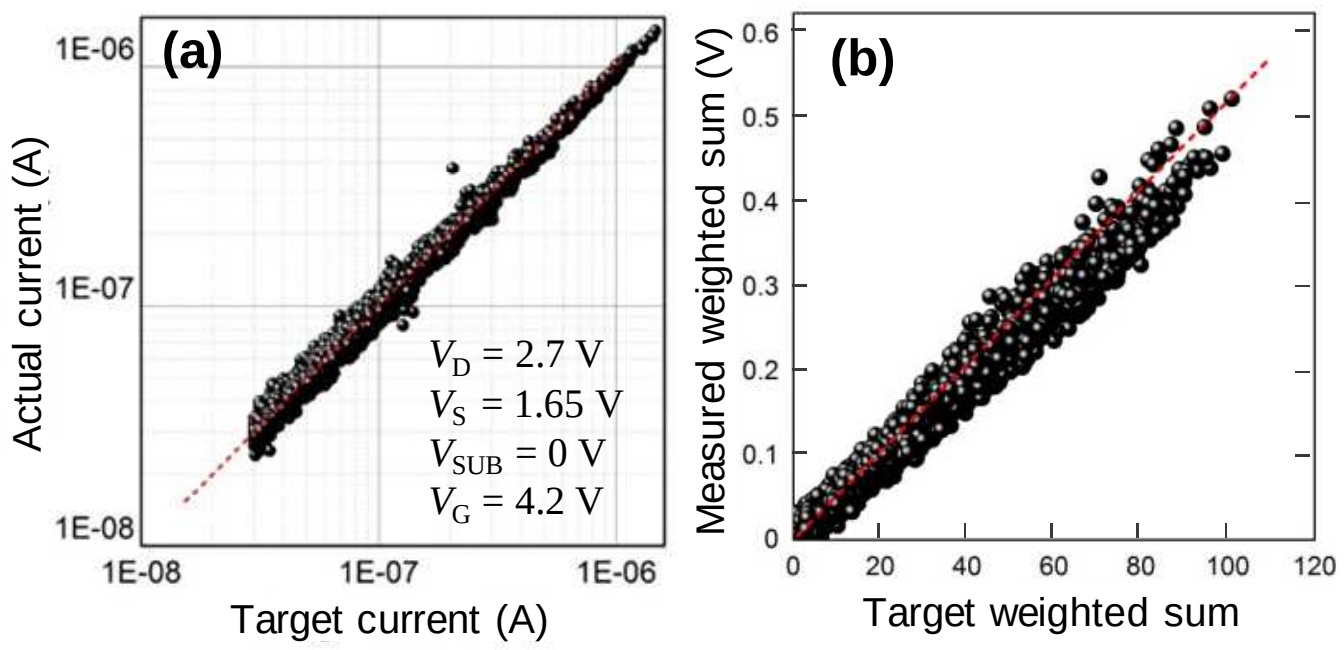

Fig. 7. (a) Comparison between the target synaptic weights (computed at the external network training) and the actual weights measured after their import, i.e. cell tuning. (b) The similar comparison for positive part of a hidden neuron output, computed for all test patterns. (Negative outputs are not shown, because they are discarded by the used activation function.) Red dashed lines are guides for the eye, corresponding to the perfect weight import.

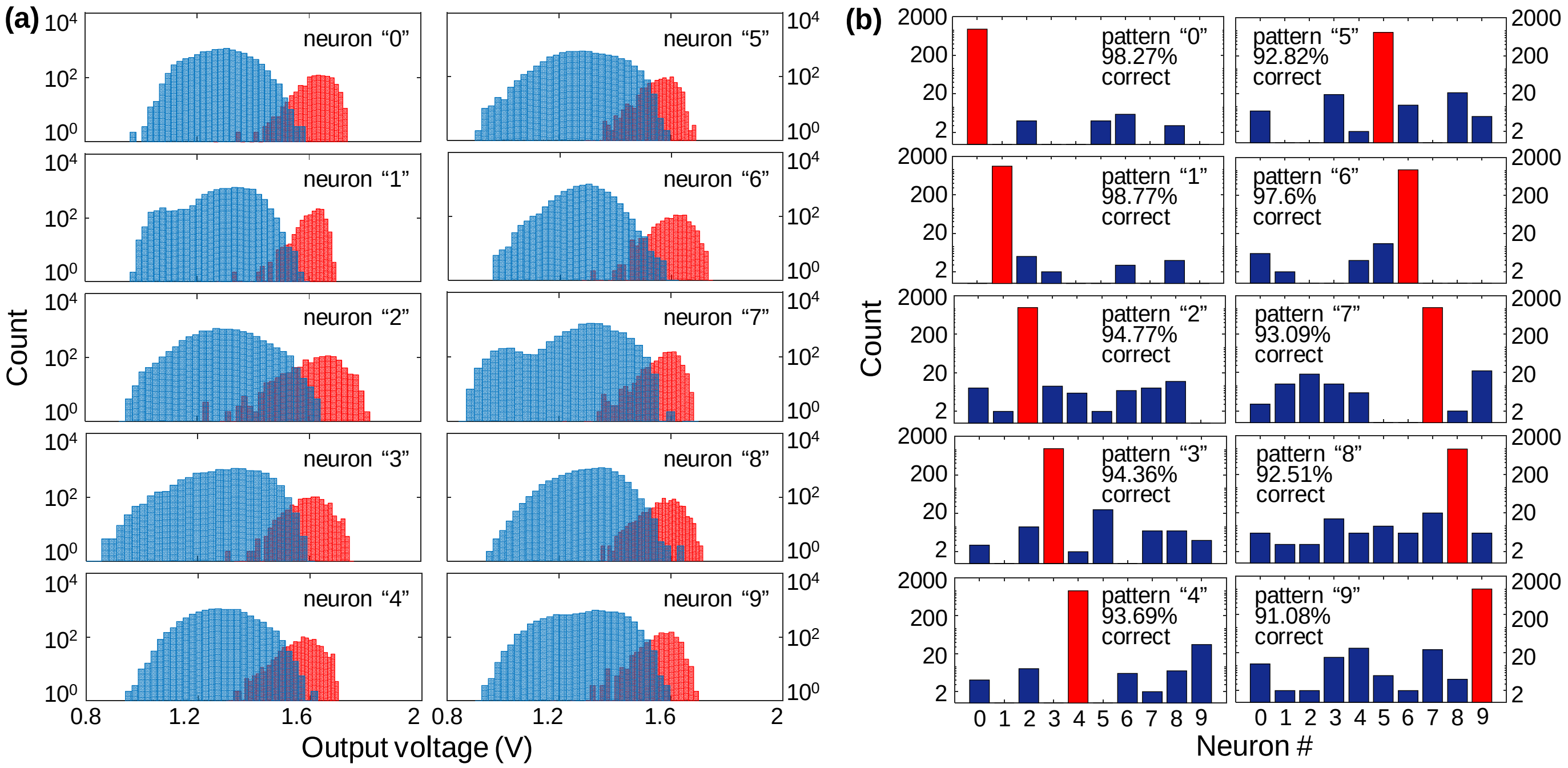

Fig. 8. Experimental results for the classification of all 10,000 MNIST test set patterns: (a) Histograms of voltages measured on each output neuron. Red bars are for the patterns corresponding to the class assigned to this particular neuron, while the blue ones are for all remaining patterns. (b) Histograms of the largest output voltages (among all output neurons) for all test patterns of each class, showing that the correct output (shown in red) always dominates. Note the log vertical scales.

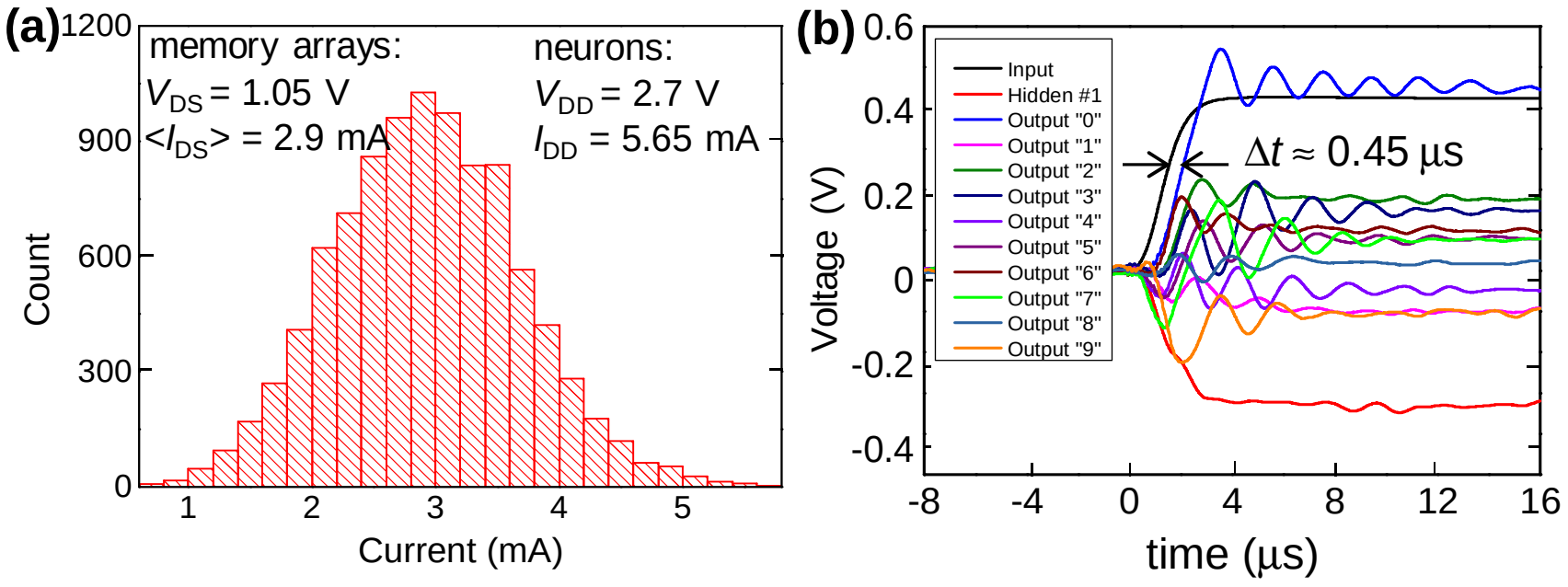

Fig. 9. (a) Histogram of the total currents flowing into the circuit, characterizing the static power consumption of both memory cell arrays, for all patterns of the MNIST test set. The inset lists the pattern-independent static current of the neurons. (b) The typical output signal dynamics after a turn-on of the voltage shifter power supply, measured simultaneously at the network input, the output of a hidden-layer neuron, and all its outputs. (The actual input voltage is ×10 larger.) The oscillatory behavior of the outputs is a result of a suboptimal phase stability design of the operational amplifiers. Before it has been improved, and the input circuit sped up, we can only claim a sub-1-µs average time delay of the network, though it is probably closer to 0.5 µs.

| AlexNet [10] single pattern classification: | Digital circuits [9] | | Mixed-signal floating-gate circuits (estimates) | | Visual cortex (estimates) |
|---|---|---|---|---|---|
| | GPU 28 nm | ASIC 65 nm | ESF1 180 nm | ESF3 55 nm | |
| time (s) | $1.5\times10^{-2}$ | $2.9\times10^{-2}$ | $\sim1\times10^{-4}$ | $\sim6\times10^{-5}$ | $\sim3\times10^{-2}$ |
| energy (J) | $1.5\times10^{-1}$ | $0.8\times10^{-2}$ | $\sim3\times10^{-7}$ | $\sim2\times10^{-7}$ | $\sim5\times10^{-8}$ |

Fig. 10. Speed and energy consumption of the signal propagation through the convolutional (dominating) part of a large deep network [10], with ~$0.65\times10^6$ neurons, at its various implementations. The estimates for floating-gate networks take into account the $55\times55$ = 3,025-step time-division multiplexing (natural for this particular network), and the experimental values of the subthreshold current slope of the cells - see, e.g., the inset in Fig. 2. The (very crude) estimate of the human visual cortex power consumption is based on the ~25 W power consumption of ~$10^{11}$ neurons of the whole brain, and a 30-ms delay of the visual cortex [11], and assumes the uniform distribution of the power over the neurons, and the same number of neurons participating in a single-pattern classification process.